# Accretion geometry of the black-hole binary Cygnus X-1 from X-ray polarimetry

M. Chauvin[1,2], H.-G. Florén[3], M. Friis[1,2], M. Jackson[1]†, T. Kamae[4,5], J. Kataoka[6], T. Kawano[7], M. Kiss[1,2], V. Mikhalev[1,2], T. Mizuno[7], N. Ohashi[7], T. Stana[1], H. Tajima[8], H. Takahashi[7]∗, N. Uchida[7], M. Pearce[1,2]

[1]Department of Physics, KTH Royal Institute of Technology, 106 91 Stockholm, Sweden.

[2]The Oskar Klein Centre for Cosmoparticle Physics, AlbaNova University Centre, 106 91 Stockholm, Sweden.

[3]Department of Astronomy, Stockholm University, 106 91 Stockholm, Sweden.

[4]Department of Physics, University of Tokyo, Tokyo 113-0033 Tokyo, Japan.

[5]SLAC/KIPAC, Stanford University, 2575 Sand Hill Road, Menlo Park, CA 94025, USA.

[6]Research Institute for Science and Engineering, Waseda University, Tokyo 169-8555, Japan.

[7]Department of Physical Science, Hiroshima University, Hiroshima 739-8526, Japan, hirotaka@astro.hiroshima-u.ac.jp.

[8]Institute for Space-Earth Environment Research, Nagoya University, Aichi 464-8601, Japan.

†Now at School of Physics and Astronomy, Cardiff University, Cardiff CF24 3AA, UK.

**Black-hole binary (BHB) systems comprise a stellar-mass black hole and a closely orbiting companion star. Matter is transferred from the companion to the black hole, forming an accretion disk, corona and jet structures. The resulting release of gravitational energy leads to emission of X-rays *(1)*. The radiation is affected by special/general relativistic effects, and can serve as a probe of the properties of the black hole and surrounding environment, if the accretion geometry is properly identified. Two competing models describe the disk-corona geometry for the hard spectral state of BHBs, based on spectral and timing measurements *(e.g., 2, 3)*. Measuring the polarization of hard X-rays reflected from the disk allows the geometry to be determined. The extent of the corona differs between the two models, affecting the strength of relativistic effects (e.g., enhancement of polarization fraction and rotation of polarization angle). Here, we report observational results on linear polarization of hard X-ray (19-181 keV) emission from a BHB, Cygnus X-1 *(e.g., 4)*, in the hard state. The low polarization fraction, <8.6% (upper limit at 90% confidence level), and the alignment of the polarization angle with the jet axis show that the dominant emission is not influenced by strong gravity. When considered together with existing spectral and timing data, our result reveals that the accretion corona is either an extended structure, or is located far from the black hole in the hard state of Cygnus X-1.**

The two competing models describe the location of the hot corona and the accretion disk in the hard state of BHBs - the 'lamp-post corona' model *(e.g., 5)* and the 'extended corona' model *(e.g., 6)*, as illustrated in Figure 1. In the lamp-post corona model, the size of the corona is small and it is located on the rotation axis of the black hole, close to the event horizon. The emission near the black hole is influenced by strong general/special relativistic effects. The extended corona model assumes a larger corona size, with the disk being truncated before reaching the innermost stable circular orbit (ISCO). In this model, primary coronal emission comes from an extended region and only a small fraction is affected by relativistic effects. Although the two models result in a comparable total flux, they predict different values for the reflection fraction (i.e., the ratio of the secondary reflection component to the primary) following the different physical parameters *(e.g., 7)*. Polarization measurements are crucial for the determination of the accretion disk geometry, providing a new probe that is independent from spectral or timing measurements *(e.g., 8)*. The polarization is characterized by the linear polarization fraction (PF) and the polarization angle (PA).

We report results from observations of Cygnus X-1 in the hard state by the PoGO+ balloon-borne polarimeter, conducted between July $12^{th}$-$18^{th}$, 2016 *(9)*. Figure 2 shows the PF and PA posterior density of PoGO+ measurements for the energy band 19–181 keV (median energy 57 keV). The *Maximum A Posteriori* estimate is PF=4.8% and PA=154°, where PA is measured from North to East (i.e., counter-clockwise on the sky). Marginalizing the posterior yields PF=$(0.0^{+5.6}_{-0.0})$% and PA=$(154\pm31)$°, where the marginalized values are obtained by the projection of the density map onto the PF and PA axis, respectively. The point-estimate and the uncertainty correspond to the peak and the region of highest posterior density containing 68.3% probability content, respectively. The 90% upper limit on the marginalized PF is 8.6% (see Supplementary Table 1). Although the PoGO+ observations lasted 6 days and covered the Cygnus X-1 orbital period of 5.6 days, no significant time variability was detected for either PF or PA.

From radio observations, the direction of the jet axis, projected onto the plane of the sky, is determined to be (158±5)° *(10, 11)*. This is consistent with our measured PA, although the uncertainty is large. Assuming that the rotation axis of the accretion disk and the radio jet are aligned, the emission can be polarized perpendicular to the plane of the accretion disk. If the primary coronal emission is polarized, our measured PA can be reproduced by Compton up-scattering on a vertically extended hot corona. Even if the primary emission is not polarized, the Compton reflected secondary component can be polarized. To produce polarization perpendicular to the disk surface, the primary emission from the hot corona must be located far above the black hole (i.e., effects of strong gravity are negligible), irradiating the outer part of the accretion disk at a grazing angle.

Assuming that the inclination of the Cygnus X-1 accretion disk is (27.1±0.8)° from that of the orbital plane *(12)* or ~40° by recent X-ray spectral analyses *(13)*, the PF is expected to be at the level of a few percent with the PA perpendicular to the accretion disk surface. This is in-line with reflection from a semi-infinite slab *(14)*, where the face-on (0°) observing geometry cancels out asymmetries (which would decrease PF) caused by the disk reflection. In the lamp-post corona case, the reflection fraction is large and the secondary reflected emission originates very close to the black hole. Doppler effects will lead to emission asymmetries (see Figure 1), yielding a relatively large PF value of up to ~15% and a PA which is no longer perpendicular to the disk surface *(15, 16)*.

To constrain the geometry quantitatively, PoGO+ observational results are compared to simulations for the lamp-post corona model *(15, 16)* and the spherically extended corona model

*(17)*. In the case of the lamp-post corona model, the polarization properties for different lamp-post corona heights are simulated for different accretion disk inclination angles and two values of black-hole spin (Schwarzschild case and extreme Kerr case). The lamp-post corona height is expressed in units of gravitational radii, $R_g = GM/c^2$, where $G$ is the gravitational constant, $M$ is the black-hole mass and $c$ is the speed of light. The spin is defined as $a = Jc/GM^2$, where $J$ is the angular momentum of the black hole (for the Schwarzschild case, $a=0$ and for the extreme Kerr case, $a=1$). The primary coronal emission is assumed to be unpolarized. If the inclination angle of the accretion disk is ~30° and the corona height is several $R_g$, the simulation predicts a relatively large PF value of ~15% (see Supplementary Figures 7, 8, 9, 10 and 11 from *(15, 16)*), due to the large reflection fraction and strong Doppler effects. Figure 3 shows the excluded parameter space (disk inclination and lamp-post corona height) resulting from the PoGO+ measurements. In *(3)*, a simple lamp-post corona scenario is applied to the X-ray spectrum of Cygnus X-1, resulting in an estimated reflection fraction ~1, $a=0.97^{+0.014}_{-0.02}$, a corona height of 5-7 $R_g$, and an inclination of $(23.7^{+6.7}_{-5.4})°$. Figure 3 shows the parameter comparison obtained for this case and for our polarization measurements. While the corona height is estimated from the spectrum to be close to the black hole, the lower PF requires a large corona height, >~50 $R_g$ and >~30 $R_g$ for the inclination angles of 27° and 40°, respectively. In the Schwarzschild case, if the inclination angle is 40°, the corona height 2-3 $R_g$ is also allowed. However, the ISCO of the disk for $a=0$ reaches only down to 6 $R_g$, resulting in faint reflected emission. We therefore conclude that the simple lamp-post corona model, which assumes a strong reflection component enhanced by strong relativistic effects, is unlikely in the hard state of Cygnus X-1.

Although the simple lamp-post corona model requires a significantly larger corona height than that from spectral/timing estimations, it may still be possible to explain the observed polarization behavior if the corona is outflowing (e.g., associated to a jet) *(18)*. The disk illumination by the corona would then be strongly affected by Doppler boosting to cancel out the effects from strong gravity. This would result in a lower reflection fraction, whereby reflection from the outer disk with grazing incidence would dominate.

We also compare our results with simulations performed for the extended corona model *(17)*. Soft X-rays from the cool part of the accretion disk are Compton up-scattered in the hot corona. The model is parameterized for several inclination angles (45°, 60° and 75°), coronal radii (equivalent to the truncated part of the cool region of the disk) and black-hole spins. Relativistic effects are smeared due to the extended corona geometry and there-by suppressed. A low PF of 2-5% is predicted for a corona radius (accretion disk inner radius) of 15 $R_g$, along with a PA perpendicular to the disk surface, which follows the expectation from analytical calculations for a semi-infinite slab *(14)* as mentioned previously. For the face-on inclination of 30˚-40° in the case of Cygnus X-1, the PF value can be even smaller than that at 45° for the same PA. This agrees well with our results and favors the extended corona model from the new viewpoint of hard X-ray polarization.

Our polarization measurements indicate that the majority of X-ray emission in the hard state of Cygnus X-1 is not influenced by strong gravity. This is compatible with the extended corona model (a hot extended corona and a cool truncated disk) proposed by previous spectral/timing analyses *(e.g., 19)*. While future observations of other BHBs are required to generalize this picture and the detailed physics of the accretion is still under discussion *(e.g., 20)*, there are reports that some observational traits (e.g., the smooth luminosity connection during transitions between soft and hard states, long reflection time lag, etc.) are also naturally explained by this model *(e.g., 21, 22)*. The extended corona model has the following implications. (i) The transition from the hard to the soft state can be caused by the shrinking of the extended corona and subsequent extension of the truncated disk to the ISCO *(19)*. The vertical extension of the corona is also evident in magnetohydrodynamic simulations for the hard state *(e.g., 23)*. (ii) The emission in the hard state primarily originates away from the black-hole horizon and is not affected by strong relativistic effects due to the black-hole spin. The spin parameter is hence difficult to estimate in the hard state by using the secondary reflection emission (e.g., the iron line). It is noted, however, that there are reports of spectral analyses with strong and broad iron lines in the hard state *(e.g., 24)*. For BHBs, the spin can instead be measured when the inner radius of the accretion disk reaches ISCO in the soft state *(1)*. For active galactic nuclei (AGNs), however, disk emission in the ultraviolet band is heavily absorbed, and a spin measurement relies solely on probing relativistic effects for the reflection component *(25)*. Hard X-ray polarization measurements, together with spectral/timing analyses, allow the reflection fraction to be determined and the geometry of the accretion disk to be elucidated, e.g. whether the optically thick part of the disk reaches ISCO.

**Correspondence:** Correspondence and requests for materials should be addressed to H.T. (email: hirotaka@astro.hiroshima-u.ac.jp).

**Acknowledgments:**

This research was supported in Sweden by The Swedish National Space Board, The Knut and Alice Wallenberg Foundation, and The Swedish Research Council. In Japan, support was provided by Japan Society for Promotion of Science and ISAS/JAXA. SSC are thanked for providing expert mission support and launch services at Esrange Space Centre. DST Control developed the PoGO+ attitude control system under the leadership of J.-E. Strömberg. Contributions from past Collaboration members and students are acknowledged. In particular, we thank M. Kole, E. Moretti, G. Olofsson and S. Rydström for their important contributions to the PoGOLite Pathfinder mission from which PoGO+ was developed.

**Author Contributions:**

M.C., H-G.F., M.F., M.J., T.Kam, J.K., T.Kaw, M.K., V.M., T.M., N.O., T.S., H.T., H.Tak., N.U. and M.P. contributed to the development of the PoGO+ mission concept and/or construction and testing of polarimeter hardware and software. Observations were conducted by M.C., H-G.F., M.F., M.K., V.M., T.S., H.Tak., N.U. and M.P. Data reduction and analysis was performed by M.C., M.F., M.K., V.M., H.Tak. and M.P. The manuscript was prepared by M.F., M.K., V.M., H.Tak. and M.P. The mission principal investigator is M.P.

**Competing Interests:** The authors declare that they have no competing financial interests.

**Author Information:**

**Affiliations:**

Department of Physics, KTH Royal Institute of Technology, 106 91 Stockholm, Sweden

Maxime Chauvin, Mette Friis, Miranda Jackson, Mózsi Kiss, Victor Mikhalev, Mark Pearce & Theodor Stana

The Oskar Klein Centre for Cosmoparticle Physics, AlbaNova University Centre, 106 91 Stockholm, Sweden

Maxime Chauvin, Mette Friis, Mózsi Kiss, Victor Mikhalev & Mark Pearce

Department of Astronomy, Stockholm University, 106 91 Stockholm, Sweden

Hans-Gustav Florén

Department of Physics, University of Tokyo, Tokyo 113-0033 Tokyo, Japan

Tuneyoshi Kamae

SLAC/KIPAC, Stanford University, 2575 Sand Hill Road, Menlo Park, CA 94025, USA

Tuneyoshi Kamae

Research Institute for Science and Engineering, Waseda University, Tokyo 169-8555, Japan

Jun Kataoka

Department of Physical Science, Hiroshima University, Hiroshima 739-8526, Japan

Takafumi Kawano, Tunefumi Mizuno, Norie Ohashi, Hiromitsu Takahashi & Nagomi Uchida

Institute for Space-Earth Environment Research, Nagoya University, Aichi 464-8601, Japan

Hiroyasu Tajima

School of Physics and Astronomy, Cardiff University, Cardiff CF24 3AA, UK

Miranda Jackson

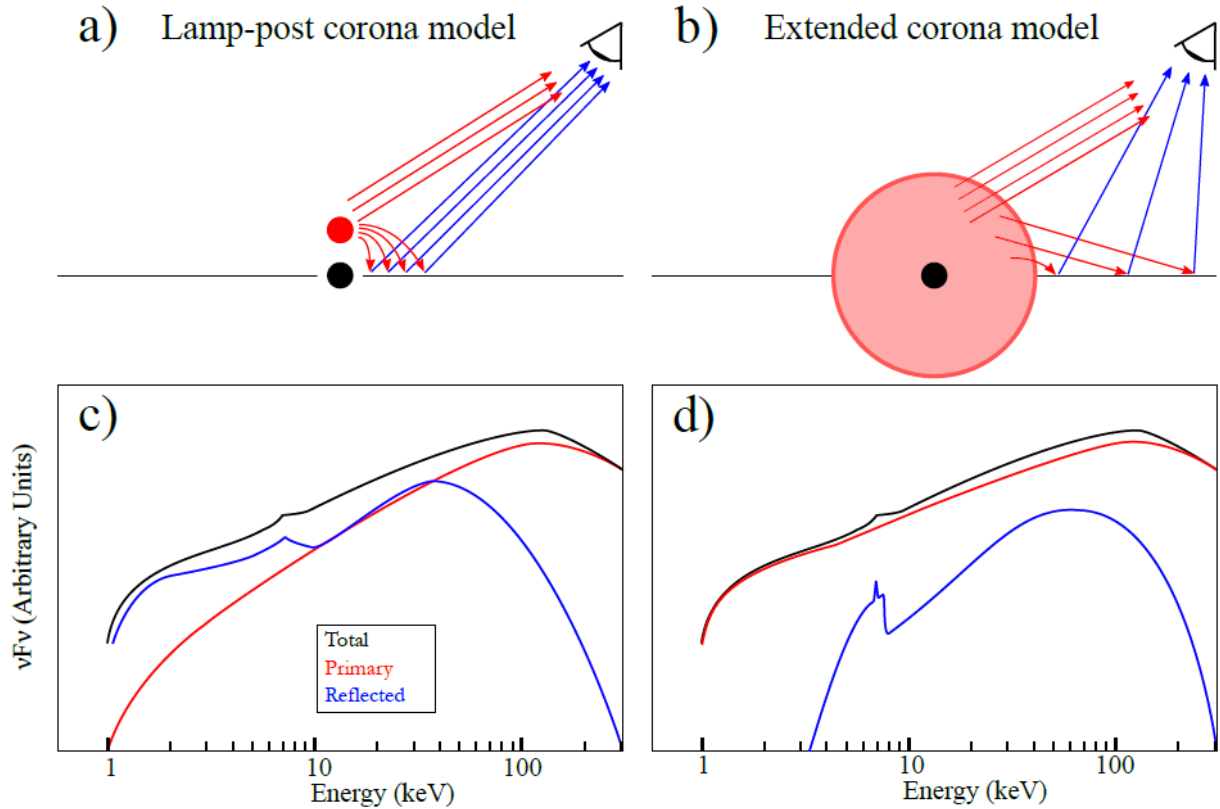

**Figure 1. Comparison of "lamp-post corona" (left) and "extended corona" (right) models describing hard X-ray emission from black hole binaries.** The geometries of the accretion disk environment are compared in the upper panels (a, b), where the black hole is shown as a black point, the cool disk is the horizontal black line, the hot corona which Compton-upscatters soft X-ray seed photons is shown in red (NB: the soft seed photons are not shown for clarity), the primary emission is shown with red arrows and the Compton reflection of the primary emission from the cool disk is shown with blue arrows. The lower panels (c, d) are a conceptual sketch of the energy spectra showing how the strong gravity of the lamp-post corona model causes an increase of the reflection fraction and a broadening of the iron lines (around 6–7 keV) due to gravitational redshift and Doppler shift. The total spectrum is shown in black. The contributions from the primary and the reflected emission are shown in red and blue, respectively.

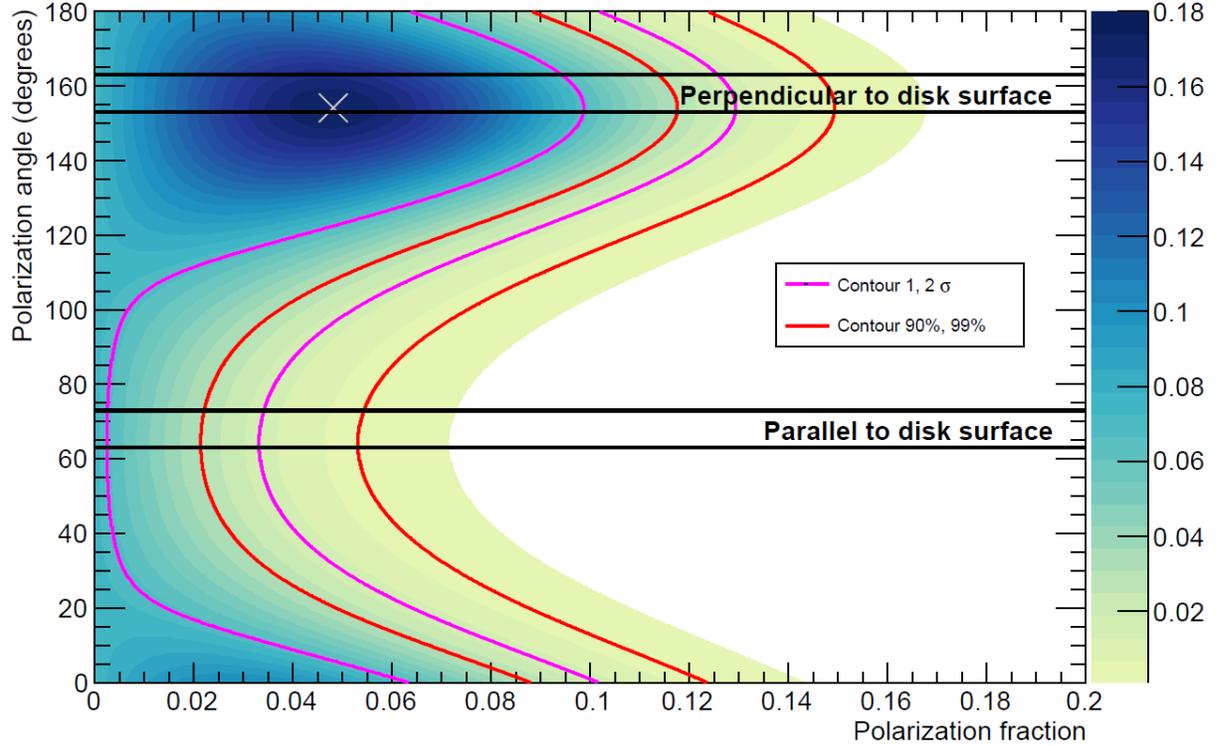

**Figure 2. Posterior density and credibility regions of PF and PA measured by PoGO+ after background subtraction.** σ is the Gaussian standard deviation. The white cross is the *Maximum A Posteriori* estimate of PF=4.8% and PA=154º. Marginalization of the posterior yields PF=(0.0$^{+5.6}_{-0.0}$)% and PA=(154±31)º (1σ confidence level, see Supplementary Table 1). The two bands show the range of the directions relative to the accretion-disk surface, which is estimated with the observed radio jet directions (±5°) *(10, 11)* being perpendicular to the surface.

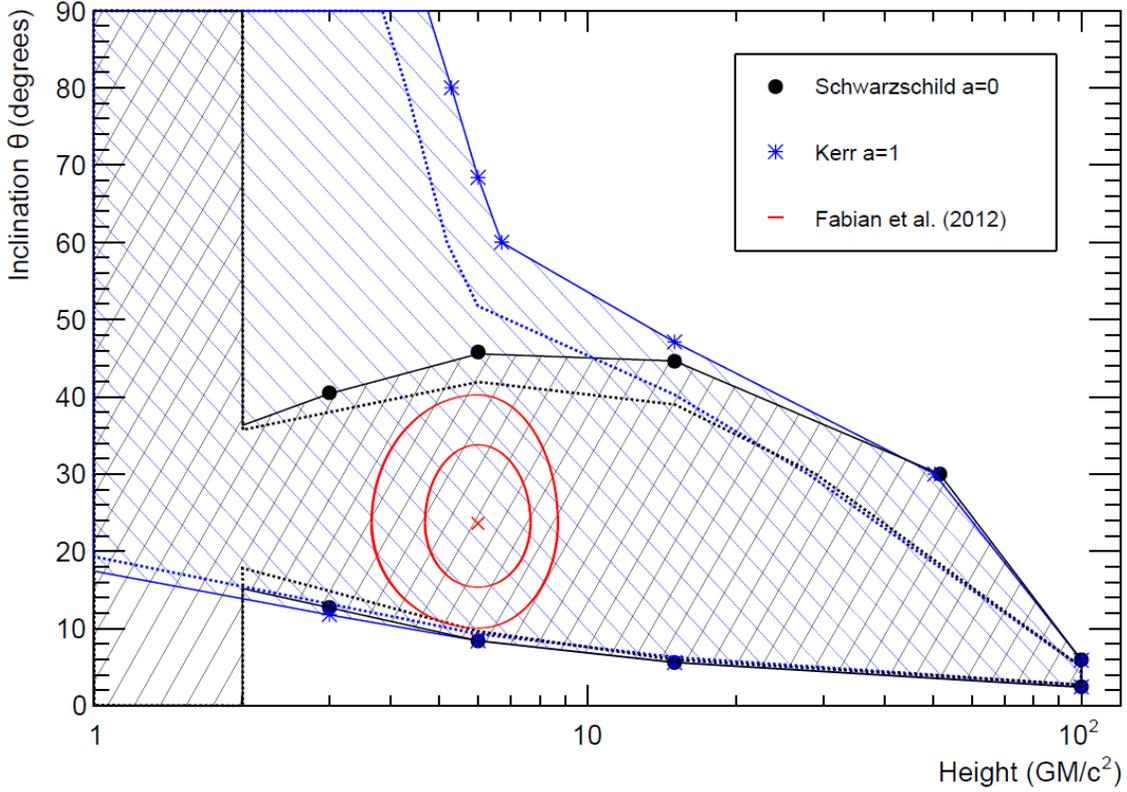

**Figure 3. Comparison of PoGO+ polarimetric and other spectral results for the lamp-post corona model.** The hashed regions show the parameter space excluded by the PoGO+ polarization measurement for the lamp-post corona model *(15, 16)*, at 90% and **2σ** credibility, as solid and dashed lines, respectively. In the case of a Schwarzschild black hole, the event horizon is located at 2 $R_g$, and the corona cannot exist below this height. For the Kerr case, both rotation directions of the accretion disk are allowed, although radio parallax measurements estimate a clockwise rotation on the sky *(12)*. The red contours are the 1- and 2-σ confidence contours derived from spectral analyses of the lamp-post corona model *(3)*, where the corona height is stated as 5-7 $R_g$, whereby the value has been taken as (6±1) $R_g$ as the 1σ confidence level. The Supplementary Information provides details on the conversion from 1D to 2D confidence intervals (see also Supplementary Figure 12).

## Methods

### X-ray emission states of BHBs

BHBs exhibit two distinct X-ray emission states referred to as the soft and hard states, which are defined by the spectral characteristics (hardness) of the emission. A thermal component is prominent in the soft state, while a radio jet is observed only in the hard state *(1)*. The hard X-ray emission consists of a primary and a secondary component. The former is well modeled by emission from the hot corona of the accretion disk. Soft X-rays originating from the inner part of the accretion disk are Compton up-scattered, forming hard X-rays within the corona. The secondary component contributes significantly around 10–200 keV, and is described as the Compton reflection of the primary emission off cool (i.e. optically thick) parts of the disk.

### Cygnus X-1

Cygnus X-1 is a high-mass X-ray binary system, comprising the first generally accepted black hole and a massive supergiant companion star *(e.g., 4)*. It is persistently bright in X-rays and is predominantly in the hard state. Due to the proximity of Cygnus X-1 to Earth ($1.86^{+0.12}_{-0.11}$ kpc) *(12)* and its brightness, it has been extensively studied and used as a prototype system for other black hole environments, notably the super-massive black holes ($10^6$-$10^{10}$ $M_\odot$) associated with AGNs *(e.g., 26)*. Observations have been conducted at energies ranging from radio (where jets are seen) *(e.g., 10, 11)*, to gamma-rays *(e.g., 27, 28)*. The black-hole mass is estimated as (15±1) $M_\odot$ *(29, 30)* from observations of the orbital motion of the companion.

### PoGO+

PoGO+ is a balloon-borne hard X-ray polarimeter mission *(9)* conducted as a collaboration between Sweden and Japan. The polarization is determined by measuring the modulation of azimuthal Compton scattering angles for incident X-rays registered in a segmented detector system. The Klein-Nishina differential cross-section dictates that X-rays are more likely to scatter in the direction perpendicular to the polarization vector. The polarimeter comprises an array of 61 hexagonal plastic scintillators (EJ-204, length of 12 cm, hexagon side length of 2.8 cm), where "2-hit events" (Compton-photoabsorption or Compton-Compton) are detected in temporal coincidence. A comprehensive anticoincidence system surrounds the plastic target. PoGO+ is a non-imaging polarimeter, and the field-of-view is restricted to ~$2° \times 2°$ by passive collimators made of Pb-Sn-Cu (67.5 cm long). During observations, the polarimeter is continuously rotated $\pm 180°$ at $1°$ s$^{-1}$ to eliminate instrumental bias.

On-ground calibration measurements show that the polarimeter does not introduce any systematic polarization (PF=$(0.10 \pm 0.12)$%) for non-polarized inputs from within the field-of-view *(9)*. As a result, the measurement of PA is independent from simulations. To derive the PF value from the measured modulation, we used laboratory measurements to correct for the instrument response and accounted for atmospheric absorption using the Geant4 toolkit *(31)*.

**PoGO+ observations of Cygnus X-1**

PoGO+ observed Cygnus X-1 between July 12$^{th}$-18$^{th}$, 2016 *(9)*. Details of the flight are provided in the Supplementary Information. Measurements are conducted in the energy range 19-181 keV (see Supplementary Figure 1 and 2), which matches the Cygnus X-1 reflection spectral feature. During observations, we monitored the Cygnus X-1 light curves reported by the MAXI instrument on the International Space Station *(32)* and the BAT instrument on the Swift satellite *(33)*. Cygnus X-1 was in the typical hard state during PoGO+ observations. The flux (20-180 keV) is estimated as $2.6\times10^{-8}$ erg s$^{-1}$ cm$^{-2}$ from a previous Suzaku *(34)* observation having similar MAXI and Swift fluxes.

From PoGO+ data, 821113 "2-hit events" were selected from 123.5 ks of Cygnus X-1 observations. Possible azimuthal anisotropy in the background can produce a fake polarization signal. To address this, interspersed observations, totaling 109.0 ks, were conducted for fields 5° to the East and to the West of the source. Transition between Cygnus X-1 and background fields occurred every ~15 minutes, in order to track temporal behavior of the background. The signal-to-background ratio is 0.1554±0.0004 (see Supplementary Figure 3). The obtained modulation curves for on-source, off-source and their subtraction are shown in Supplementary Figure 4, 5 and 6, respectively. Polarization parameters are derived using unbinned and background-subtracted Stokes parameters. The observational and analysis methods are the same as those utilized for Crab observations conducted during the same PoGO+ flight *(35)*.

**Previous polarization observations of Cygnus X-1**

Polarization measurements of Cygnus X-1 have been carried out for a wide range of energies *(36)*. For X-rays, measurement are reported at 2.6 keV, 5.2 keV by OSO-8 *(37)* and 230-2000 keV by INTEGRAL *(38, 39)*. These energy bands are not suitable to probe the reflected emission component.

There are reports of highly-polarized (PF >75%) emission with PA=40º in the energy band >230 keV by INTEGRAL *(38, 39)*. The hard X-ray emission was claimed to arise from jet synchrotron emission *(e.g., 40)*. In our work, since PoGO+ data (PA~160º in Figure 2) prefer the orthogonal component rather than the PA=40º one, we ignore a possible contribution from the emission studied by INTEGRAL and consider only the reflection component. We note that the polarization analysis of INTEGRAL data is complicated by the lack of pre-launch studies of the polarimetric instrument response.

**Comparisons with model simulations**

The model simulations of the lamp-post and extended corona models cover only part of the PoGO+ energy range (discrete energy levels in the range 20–50 keV in *(15, 16)* and <100 keV in *(17)*, as compared to the PoGO+ energy range of 19-181 keV with a median of 57 keV). At the higher energies measured by PoGO+, the lower opacity of the cool part of the accretion disk could result in a larger PF. However, this effect is expected to be compensated by the energy dependence of scattering interactions *(41)*. The value of PA is not expected to be significantly affected and therefore our conclusions stand.

The current simulation of the lamp-post corona model assumes unpolarized primary emission *(15, 16)*. Intrinsic polarization, e.g. by Compton up-scattering, may also affect the estimations, as may the size/structure/outflow characteristics of the corona. To estimate the corona radius (inner disk radius) in the extended corona model, new simulations are required. Currently, the radius dependency is reported only for the inclination angle 75° *(17)*, where PoGO+ results allow the parameter region for radii >~10 $R_g$ and exclude the region <5 $R_g$. Another corona geometry has been simulated with the same simulation scheme as the extended corona model *(17)*. The corona layer sandwiches the disk surface and the disk and corona both extend from ISCO to the outer radius. In this case, due to the large contribution far from the black hole, strong relativistic effects are obscured. The PF is predicted to be as low as 4-8% with the PA perpendicular to the disk surface, which again matches the PoGO+ measurements.

**Data availability:** The data that support the plots within this paper and other findings of this study are available from the corresponding author upon reasonable request.

# Supplementary Materials for

**Accretion geometry of the black-hole binary Cygnus X-1 from X-ray polarimetry**

M. Chauvin[1,2], H.-G. Florén[3], M. Friis[1,2], M. Jackson[1]†, T. Kamae[4,5], J. Kataoka[6], T. Kawano[7], M. Kiss[1,2], V. Mikhalev[1,2], T. Mizuno[7], N. Ohashi[7], T. Stana[1], H. Tajima[8], H. Takahashi[7]*, N. Uchida[7], M. Pearce[1,2]

[1]Department of Physics, KTH Royal Institute of Technology, 106 91 Stockholm, Sweden.

[2]The Oskar Klein Centre for Cosmoparticle Physics, AlbaNova University Centre, 106 91 Stockholm, Sweden.

[3]Department of Astronomy, Stockholm University, 106 91 Stockholm, Sweden.

[4]Department of Physics, University of Tokyo, Tokyo 113-0033, Japan.

[5]SLAC/KIPAC, Stanford University, 2575 Sand Hill Road, Menlo Park, CA 94025, USA

[6]Research Institute for Science and Engineering, Waseda University, Tokyo 169-8555, Japan.

[7]Department of Physical Science, Hiroshima University, Hiroshima 739-8526, Japan.

[8]Institute for Space-Earth Environment Research, Nagoya University, Aichi 464-8601, Japan.

*Correspondence to: hirotaka@astro.hiroshima-u.ac.jp.

†Now at School of Physics and Astronomy, Cardiff University, Cardiff CF24 3AA, UK.

**Introduction**

The measurements presented in this paper were conducted with the balloon-borne hard X-ray polarimeter PoGO+. During the same balloon flight, the polarization of emission from the Crab was also measured, as described in *(1, 2)*. The Supplementary Information of the Crab paper *(1)* provides a detailed discussion of the mathematical framework used for measuring polarization, a description of the PoGO+ polarimeter and ancillary systems, a review of the balloon flight, the data reduction method, polarization analysis, and studies of background anisotropy. The reader is advised to read the Crab Supplementary Information *(1)* as background to this document, which details aspects of the Cygnus X-1 analysis that differ from those presented in the Crab paper.

**Polarimetric performance**

The effective area of the polarimeter for Cygnus X-1 observations is shown in Supplementary Figure 1. Data are derived from Geant4 simulations *(3)* using the atmospheric column density profile measured during Cygnus X-1 observations. Supplementary Figure 2 shows the simulated count rate as a function of energy for the polarization analysis. The measured signal rate is ~1 Hz during the observations. The spectral shape is estimated from a previous Suzaku observation *(4)* which has a similar flux and hardness ratio as measured by the MAXI instrument on the International Space Station *(5)* and the BAT instrument on the Swift satellite *(6)*. The MAXI and BAT observations were contemporaneous with the PoGO+ measurements. The resulting spectral index is -1.3, with a cut-off energy at 146 keV. For these values, simulations determine that the energy range of PoGO+ for Cygnus X-1 observations is 19–181 keV (median energy is 57 keV). For this spectrum, the simulated $M_{100}$, the response to a 100% polarized beam, is (44.09±0.76)%. During on-ground calibration studies prior to the flight *(7)*, a polarization fraction of (0.10±0.12)% was measured for an unpolarized beam, thereby showing that no systematic polarization signal is induced.

**Observations and data reduction**

During Cygnus X-1 observations, an optical star tracker camera is used as the primary pointing system, augmented by an Inertial Measurement Unit (IMU) which corrects for high frequency disturbances coming from the balloon rigging, bearing friction and polarimeter cabling. The star camera tracks on a nearby reference star, HIP 98110. PoGO+ employs collimated detectors resulting in a field-of-view of ~2° × 2°. A pointing precision of 0.1° is required in order to not degrade the performance due to collimator shadowing effects.

Cygnus X-1 was observed on each day of the flight between July $12^{th}$ – $18^{th}$, 2016, for a total of 6 observations with an average atmospheric column density of 6.5 g/cm$^2$. Observations are performed by interspersing on-source and off-source measurements in ~15 minute intervals. This allows for background subtraction and elimination of possible fake polarization events. After accounting for the dead-time of the data acquisition system, the total observation time was 123.5 ks on Cygnus X-1 and 109.6 ks for the background fields. Selections developed during laboratory tests of the polarimeter were applied to flight data, as described in *(7)*. Supplementary Figure 3 shows the count rate for interspersed on-source and off-source measurements. Before background subtraction, a Minimum Detectable Polarization of 8% is achieved for the Cygnus X-1 observations with a signal-to-background ratio of 0.1554±0.0004. The observation comprises 821823 2-hit events from the Cygnus X-1 observations and 631203 2-hit events from the background fields. Such "2-hit events" refer to two time-coincident hits in separate plastic scintillator detectors. These events allow for the calculation of a scattering angle for incident X-rays and the subsequent determination of polarization properties from the azimuthal angle dependence of the Compton scattering cross-section.

**Polarization analysis**

Supplementary Figures 4, 5 and 6 show the azimuthal scattering angle distributions (modulation curves) for the on-source, off-source and background-subtracted on-source data. A $\chi^2$ fit is performed and the resulting modulation curves inspected for goodness-of-fit. Anisotropic background conditions can lead to distortions of the modulation curve. Although the on-source and off-source require a 360° sinusoidal component to achieve a good fit, the background subtracted curve does not. The background conditions were found to be very similar to those for the Crab observation, as described in the Supplementary Information of *(1)*.

The amplitude of the modulation, *A*, in Supplementary Figure 6 is small relative to its uncertainty, therefore conventional Stokes parameters analysis cannot be used since the assumption of Gaussian shape of the uncertainties is no longer valid as discussed in *(8)*. Using the analysis techniques described in the Supplementary Information of *(1)*, the posterior distribution for the polarization parameters is calculated and shown in Figure 2. The maximum a posteriori (MAP) estimates $(p, \psi)$, as well as the marginalized $(\Pi, \Psi)$ parameters, are summarized in Supplementary Table 1. The marginalized distribution for $\Pi$ is used to calculate an upper limit on the polarization fraction, at 90% confidence, of 8.6%.

**Lamp-post corona model parameters**

In order to produce Figure 3, two parameters (the height of the primary emission and the inclination angle) of the lamp-post corona model were considered. Supplementary Figure 7 illustrates how rejection regions are determined. Simulated predictions *(9, 10)* for different lamp-post corona heights from ISCO (Innermost Stable Circular Orbit) to $100R_g$ (where $R_g = GM/c^2$) are overlaid on the PoGO+ polarization parameter contour plots for a fixed inclination of the accretion disk at 30°, 60° and 80°. Results for both black-hole rotation directions are plotted, although this probably follows the orbital rotation, which is clockwise on the sky, as estimated from radio parallax measurements *(11)*. To exclude parameter regions, we require the PF value for both rotational directions to be excluded. Supplementary Figures 8 and 9 show the rejected regions for a Schwarzschild black hole while Supplementary Figure 10 and 11 show the rejected regions for the Kerr scenario. Finally, the results from Supplementary Figures 8–11 are combined in Figure 3. Exclusion points are connected with a straight line to form a closed exclusion region.

The red contours in Figure 3 are results from *(12)*, where spectral fits are used to constrain lamp-post corona model parameters with a simple modeling, yielding the spin $a = 0.97^{+0.014}_{-0.02}$ (the error is the 90% confidence level). The corona height and the inclination are estimated as 5–7 $R_g$ and $(23.7^{+6.7}_{-5.4})°$ with the confidence level unspecified. For Figure 3, we assume the height of (6±1) $R_g$ and these error ranges to be the 1σ confidence level. If we treat the uncertainties as the 90% confidence level, which matches the confidence level on the spin, the red contours become as shown in Supplementary Figure 12. Although more sophisticated spectral modeling of the lamp-post corona emission exists *(e.g, 13)*, recent works reported inconsistencies within the model *(14)*. We have therefore opted to use the simple model described in *(12)*. We convert these 1D height and inclination 1σ intervals to 2D contours by scaling the uncertainty with a $\chi^2$ factor of $\sqrt{2.30}$ and $\sqrt{6.18}$ for the $1\sigma$ and $2\sigma$ contours, respectively. Since the uncertainty on the inclination is asymmetric and the height is displayed on a log scale, each contour is determined by dividing it into quadrants. Each quadrant encompasses the arc of an ellipse whose semi-major and semi-minor axes correspond to the uncertainties for that quadrant.

**Supplementary Table 1. Polarization results for the Cygnus X-1 observation.** The maximum a posteriori (MAP) probability estimate is given by $(p, \psi)$ and the associated uncertainties $(\sigma_p, \sigma_\psi)$. The peaks of the marginalized posteriors are given by $(\Pi, \Psi)$. The measurement of the polarization fraction is positive-definite, $p > \Pi$ for all entries, i.e. the MAP result is always greater than the marginalized estimate of the polarization fraction. The angle is not governed by the same statistics and thus $\psi = \Psi$. Their uncertainties, however, are different, with $\sigma_\Psi > \sigma_\psi$ arising as a direct consequence of $p > \Pi$. A detailed description of the underlying mathematical formulas is given in Supplementary Information of *(1)*. The quoted error range is 1σ (68.3%) confidence level, except for the 90% upper limit.

| MAP fraction $p$ (%) | Marg. fraction $\Pi$ (%) | 90% Upper Limit $\Pi$ (%) | MAP angle $\psi$ (°) | Marg. angle $\Psi$ (°) |
|---|---|---|---|---|
| 4.81±3.73 | $0^{+5.6}_{-0}$ | 8.6 | 154±22 | 154±31 |

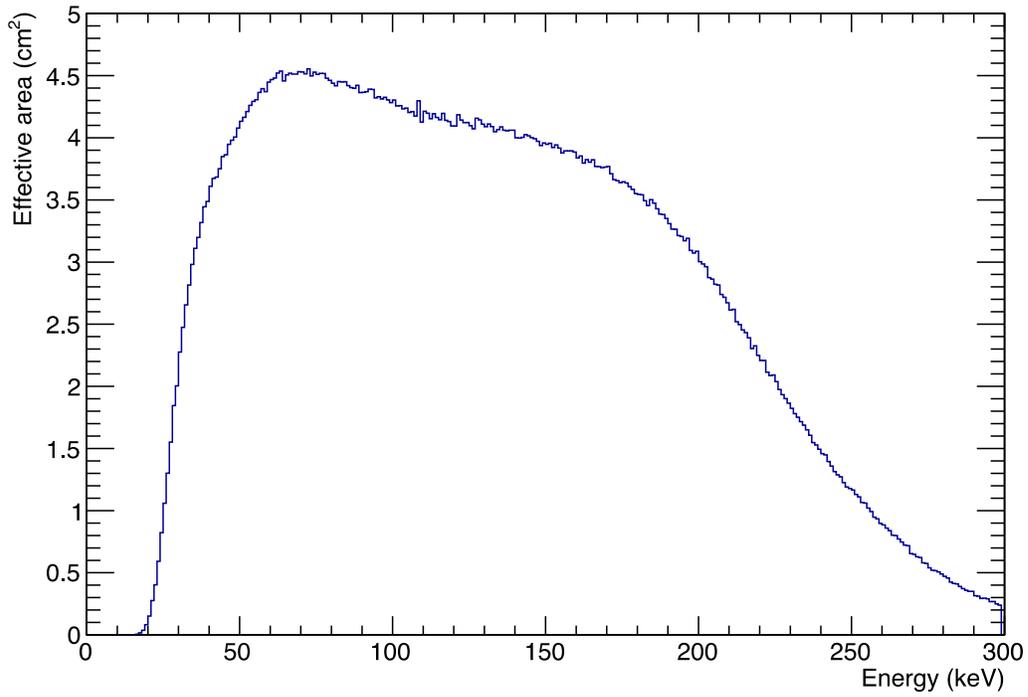

**Supplementary Figure 1. Simulated effective area as a function of energy.** The effective area is simulated using the column density measured during the flight. Events in the range 19–181 keV (as determined from Supplementary Figure 2) are used for the polarization analysis.

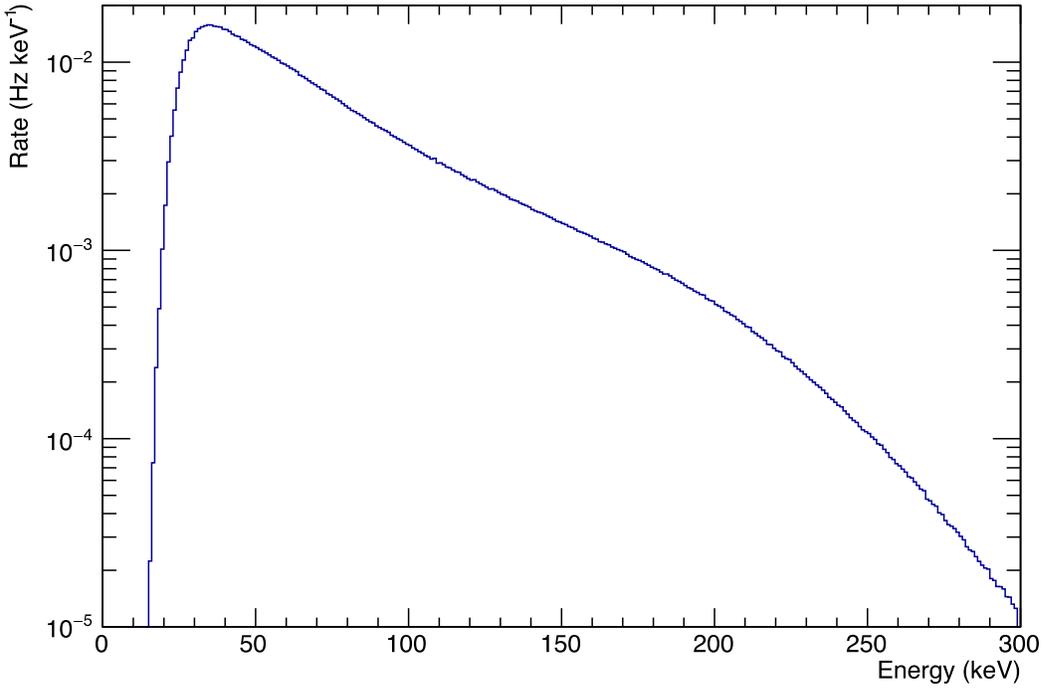

**Supplementary Figure 2. Simulated count rate as a function of energy.** The count rate is a product of the effective area shown in Supplementary Figure 1 and the Cygnus X-1 energy spectrum, as measured by the Suzaku satellite. The energy range for polarization measurements is defined as 19–181 keV. Outside these limits the count rate drops below 5% of the peak value.

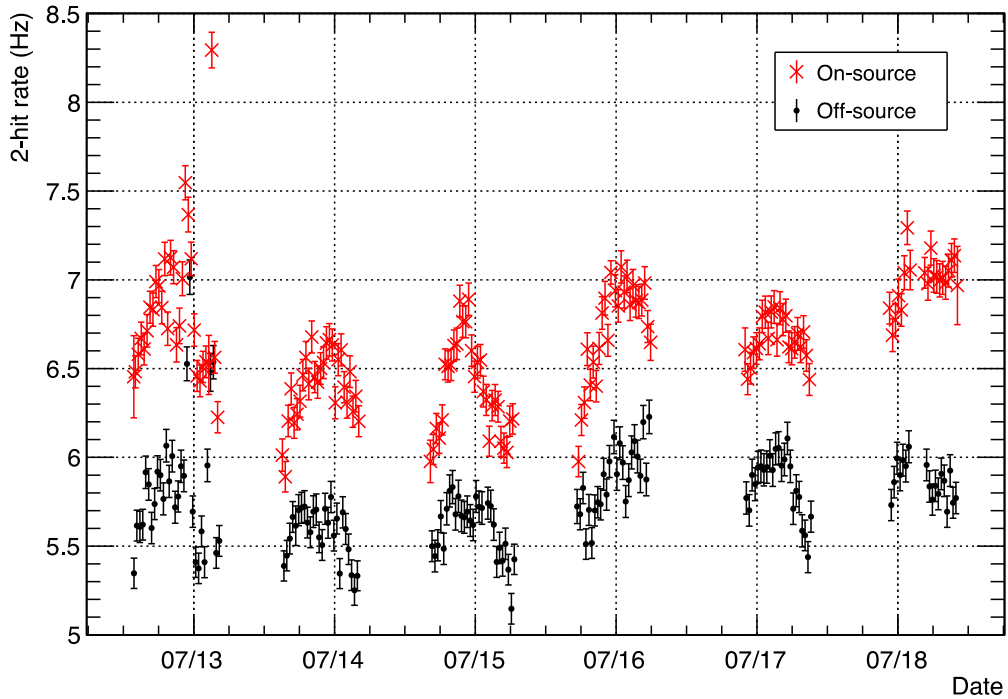

**Supplementary Figure 3. Light curve of 2-hit polarization events.** Count rates of 2-hit events used for the polarization analysis are shown for on-source (red crosses) and off-source (black dots) observations as a function of date during the 2016 flight. The error range is 1σ (68.3%) confidence level. The 2-hit events consist of two coincident hits in separate scintillator detectors, allowing for calculation of an azimuthal scattering angle and subsequent determination of polarization parameters.

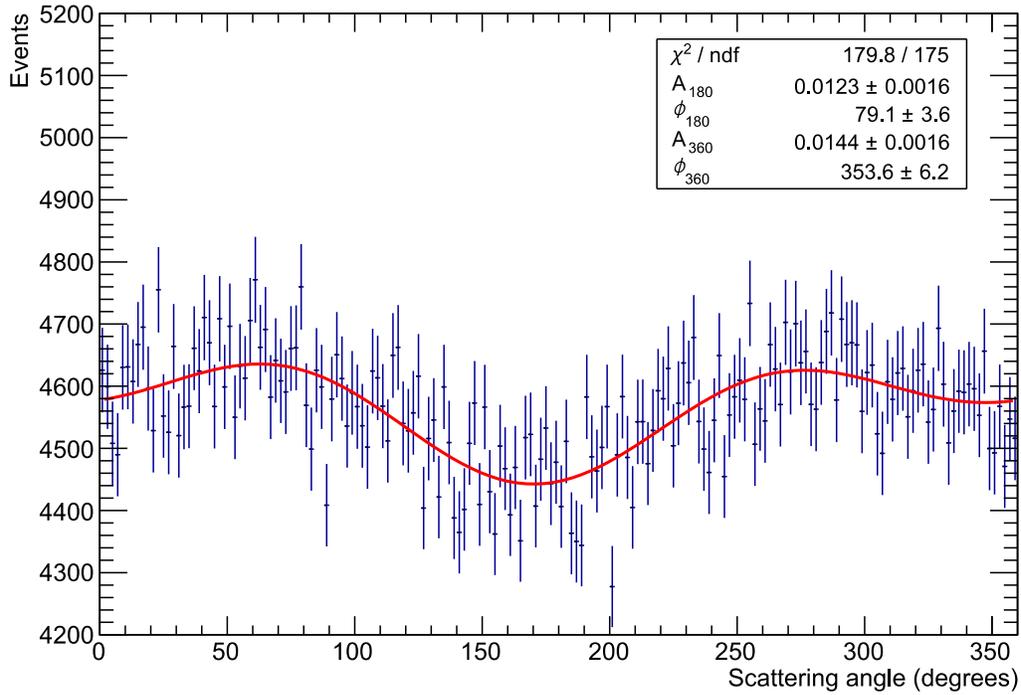

**Supplementary Figure 4. Modulation curve for on-source measurements.** Scattering angle distribution (blue points) and fitted modulation curve (red line) for the Cygnus X-1 observations. The amplitude and phase of the fitted sinusoidal curves are *A* and *ϕ*, respectively, with the index indicating the 180° and 360° components. The error ranges are 1σ (68.3%) confidence level. Both a 180° and a 360° component are needed to provide a good fit, indicating that the background incident on the polarimeter is anisotropic.

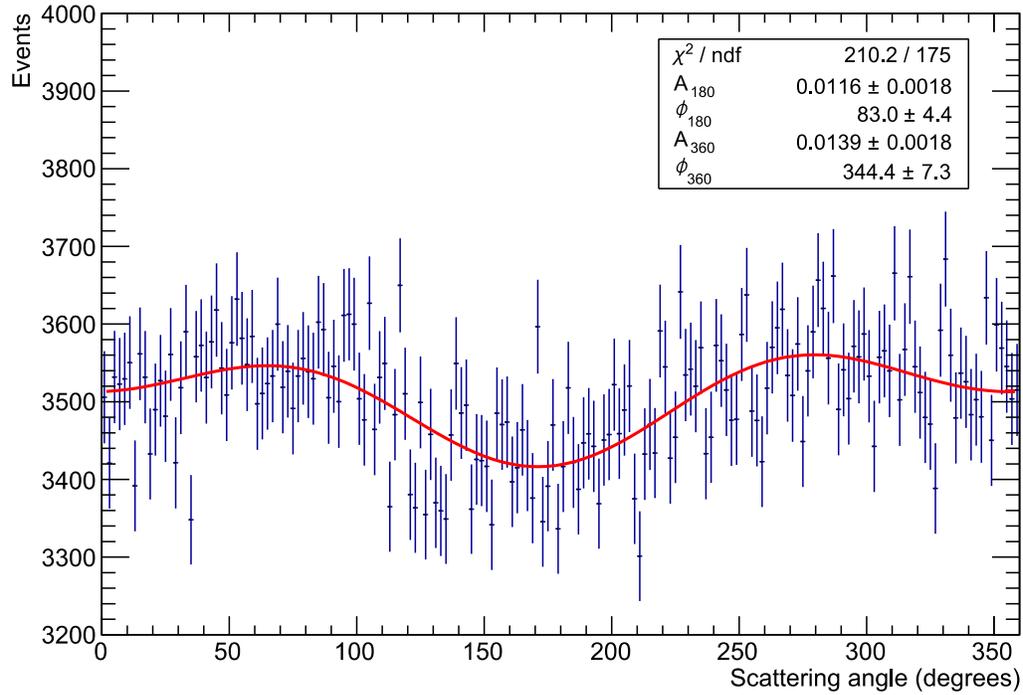

**Supplementary Figure 5. Modulation curve for off-source measurements.** Scattering angle distribution (blue points) and fitted modulation curve (red line) for the Cygnus X-1 background (off-source) observations. The amplitude and phase of the fitted sinusoidal curves are $A$ and $\phi$, respectively, with the index indicating the 180° and 360° components. The error ranges are 1σ (68.3%) confidence level. Both a 180° and a 360° component are needed to provide a good fit.

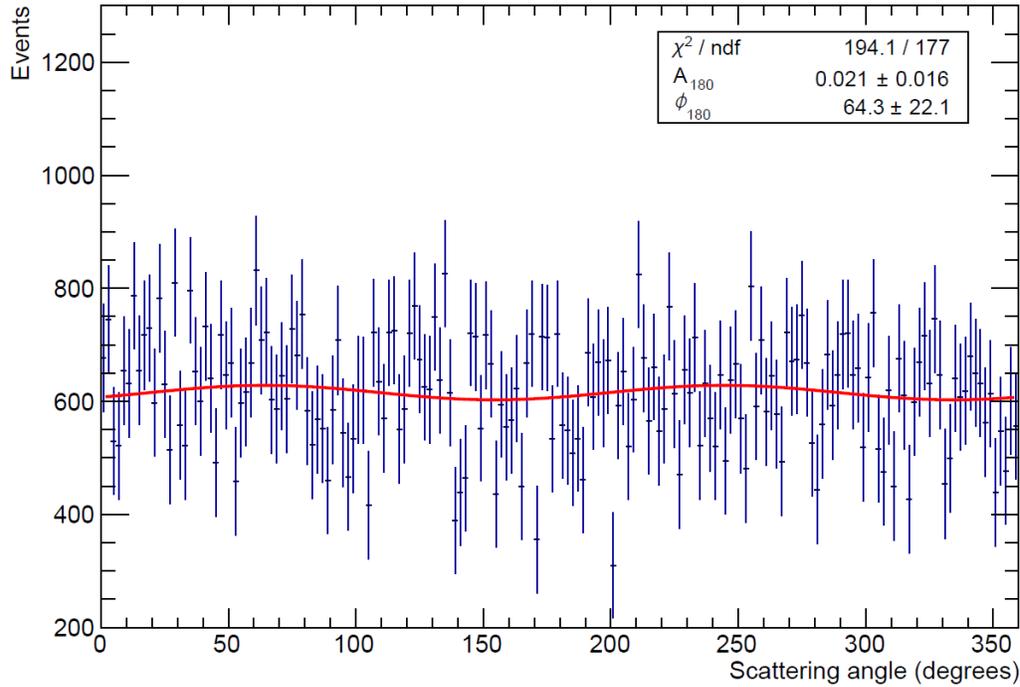

**Supplementary Figure 6. Modulation curve for the background-subtracted measurements.** Scattering angle distribution (blue points) and fitted modulation curve (red line) for the Cygnus X-1 observation following background subtraction. The amplitude and phase of the fitted sinusoidal curve are $A$ and $\phi$, respectively, with the index indicating the 180° component. The error ranges are 1$\sigma$ (68.3%) confidence level. As indicated by the chi-square value, a 360° component is no longer needed to provide a good fit.

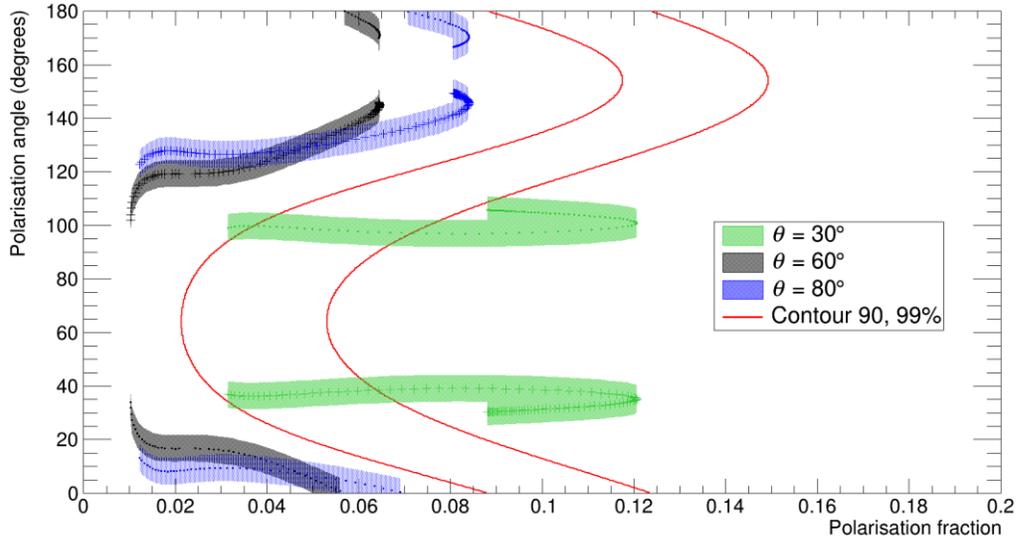

**Supplementary Figure 7. Comparison of lamp height to PoGO+ results.** The polarization parameters resulting from lamp heights from ISCO to $100R_g$ are plotted assuming a Schwarzschild black hole and a fixed inclination, $\theta$, of the accretion disk *(9, 10)*. The cross symbol corresponds to rotation of the black hole (and hence the accretion disk) counter-clockwise on the sky, while the dots denote data points corresponding to the clockwise direction. As indicated by the colored bands, we assume an uncertainty of ±5° on the simulated polarization angle, corresponding to the uncertainty in the jet direction projected onto the sky as determined by *(15, 16)*. The credibility regions from PoGO+ measurements are shown as red lines.

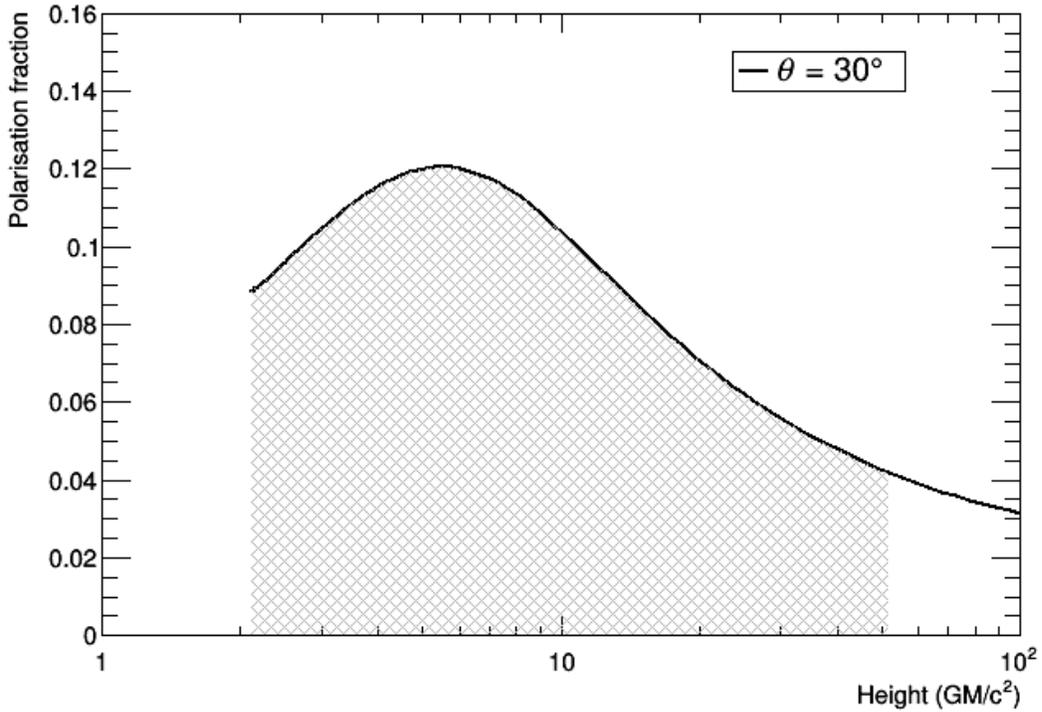

**Supplementary Figure 8. Rejection region of lamp-post corona height for assumed inclination.** Polarization fraction as a function of lamp-post corona height for a fixed accretion disk inclination. The curve comes from lamp-post corona model simulations showing the predicted polarization fraction in the range 20-50 keV *(9, 10)*. The shaded region corresponds to the rejected parameter space for a chosen inclination of 30° (no heights at 60° and 80° are rejected) and assuming a Schwarzschild black hole, as shown in Supplementary Figure 7.

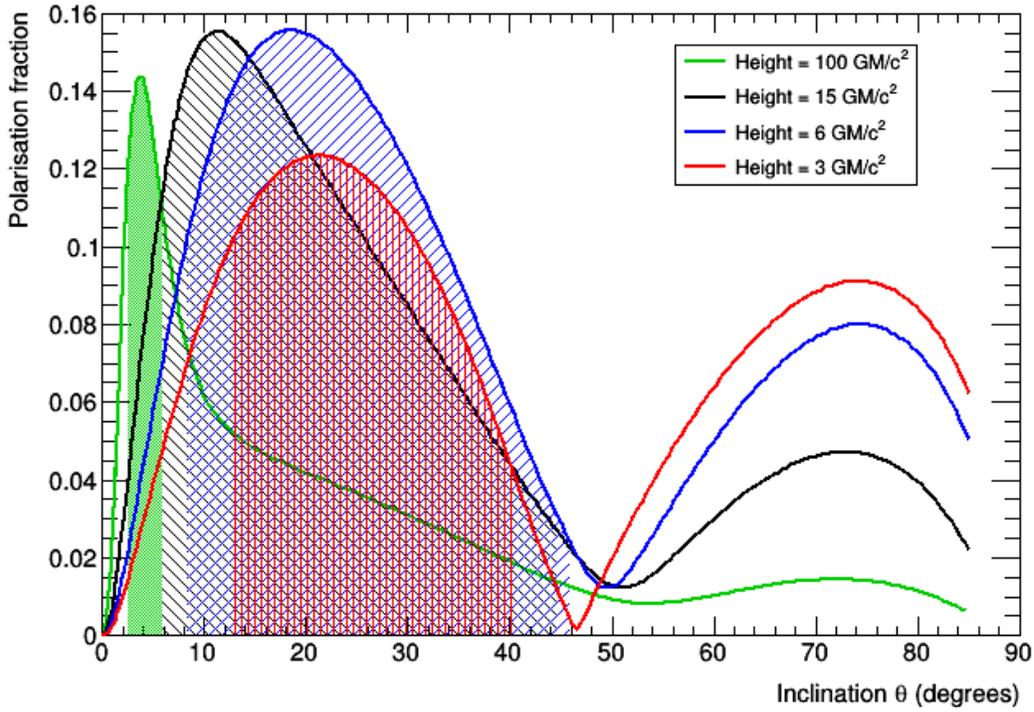

**Supplementary Figure 9. Rejection regions of lamp-post model inclination for assumed lamp heights.** Same as Supplementary Figure 8, but as a function of accretion disk inclination for fixed lamp-post corona heights. Rejection regions are shown for four different lamp heights.

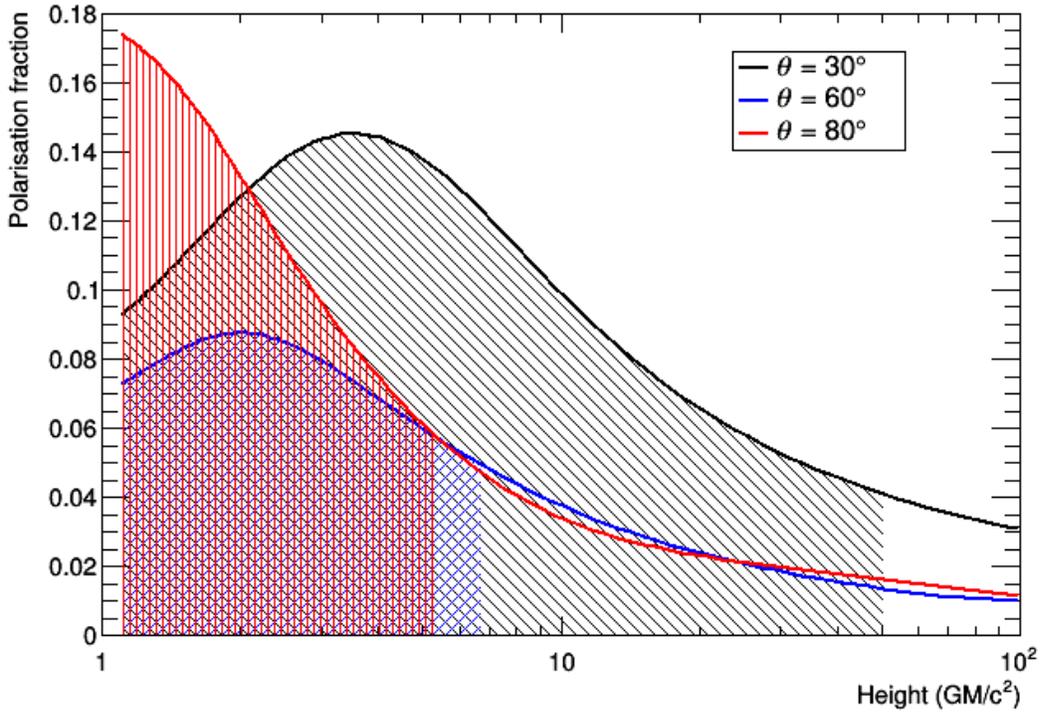

**Supplementary Figure 10. Rejection region of lamp-post corona height for assumed inclination.** Same as Supplementary Figure 8, but assuming a Kerr black hole.

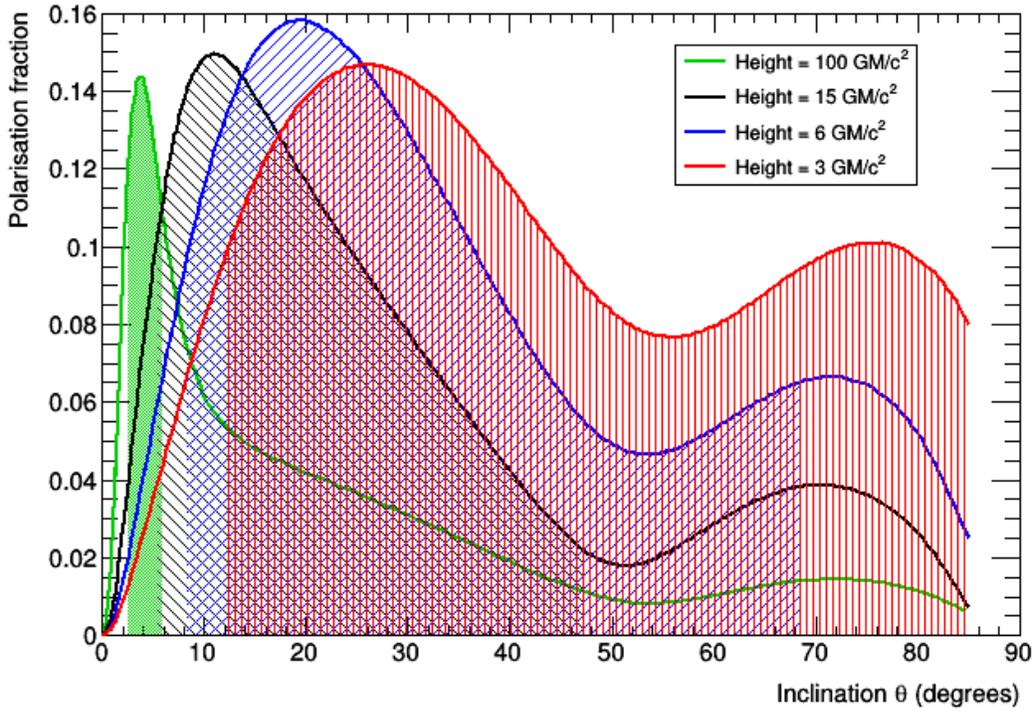

**Supplementary Figure 11. Rejection region of lamp-post model inclination for assumed lamp heights.** Same as Supplementary Figure 9, but assuming a Kerr black hole.

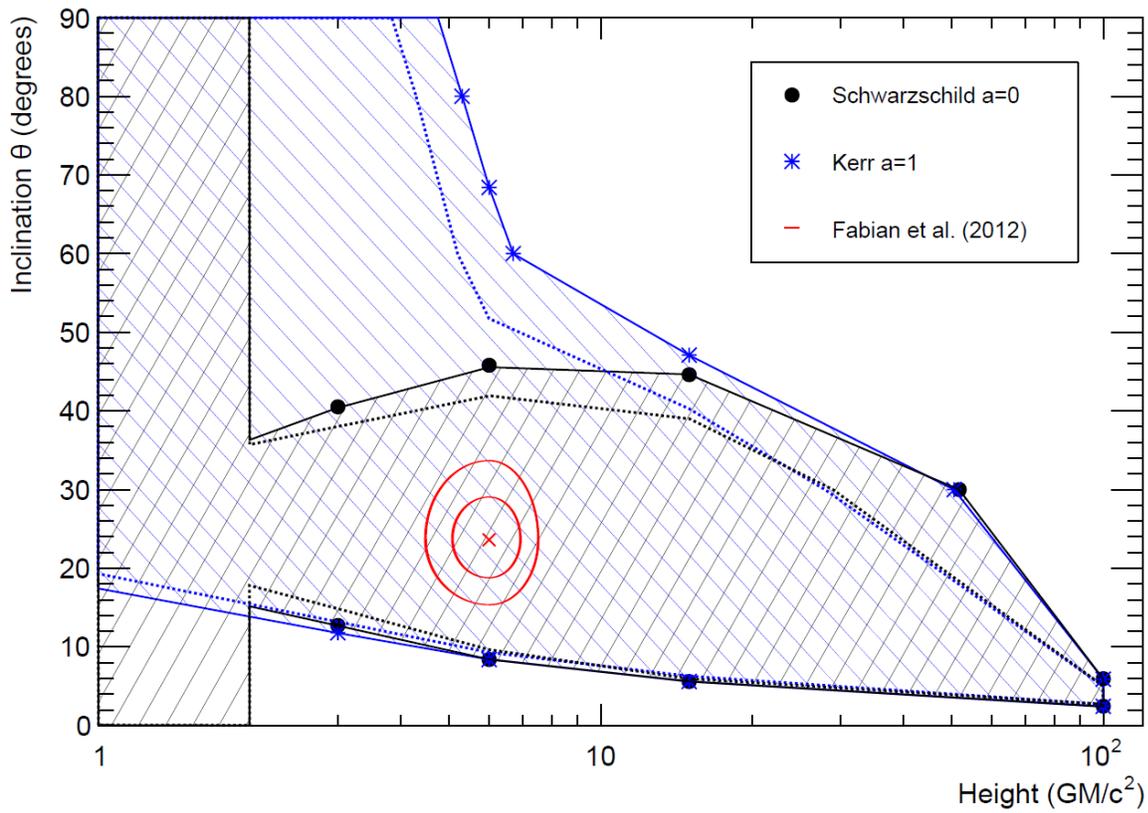

**Supplementary Figure 12. Comparison of PoGO+ polarimetric and other spectral results for the lamp-post corona model.** Same as Figure 3, but assuming the error ranges estimated by the spectral analysis *(12)* to be the 90% (not 1σ) confidence level for the corona height and the inclination.